\documentclass{emulateapj} 
\slugcomment{Astrophys. J., Accepted, April 11 2008}

\def\rms{\mathrm{s}}

\def\be{\begin{equation}}
\def\ee{\end{equation}}
\def\ts{t_\rms}

\begin{document}
\title
{Towards planetesimals: dense chondrule clumps \\ 
in the protoplanetary nebula}
\author {Jeffrey N. Cuzzi\altaffilmark{1,4}, Robert C. Hogan\altaffilmark{2}, and Karim Shariff\altaffilmark{3}}
\altaffiltext{1}{Space Science Division, Ames Research Center, Moffett Field, CA 94035, USA}
\altaffiltext{2}{BAER inc., Sonoma, CA} 
\altaffiltext{3}{NASA Advanced Supercomputing Division, Ames Research Center} 
\altaffiltext{4}{To whom correspondence should be addressed; E-mail: 
jeffrey.cuzzi@nasa.gov}
\begin{abstract}
We outline a scenario which traces a direct path from freely-floating nebula particles to the first 10-100km-sized bodies in the terrestrial planet region, producing planetesimals which have properties matching those of primitive meteorite parent bodies. We call this {\it primary accretion}. The scenario draws on elements of previous work, and introduces a new critical threshold for planetesimal formation. We presume the nebula to be weakly turbulent, which leads to dense concentrations of aerodynamically size-sorted particles having properties like those observed in chondrites. The fractional volume of the nebula occupied by these dense zones or clumps obeys a probability distribution as a function of their density, and the densest concentrations have particle mass density 100 times that of the gas. However, even these densest clumps are prevented by gas pressure from undergoing gravitational instability in the traditional sense (on a dynamical timescale). While in this state of arrested development, they are susceptible to disruption by the ram pressure of the differentially orbiting nebula gas. However, self-gravity can preserve sufficiently large and dense clumps from ram pressure disruption, allowing their entrained particles to sediment gently but inexorably towards their centers, producing 10-100 km ``sandpile" planetesimals. Localized radial pressure fluctuations in the nebula, and interactions between differentially moving dense clumps, will also play a role that must be allowed for in future studies. The scenario is readily extended from meteorite parent bodies to primary accretion throughout the solar system.
\end{abstract}

\section{Introduction} 
There is no currently accepted scenario for the
formation of the parent bodies of primitive meteorites which accounts for the 
most obvious of their properties. These properties 
(reviewed by Scott and Krot 2005 and discussed 
in more detail in section 2.1) include (a) dominance by aerodynamically well-sorted mineral particles of sub-mm size; (b)
class-to-class variation in well-defined physical, chemical, and isotopic
properties; (c) a spread of 1 Myr or so between the formation times of the
oldest and youngest objects found in the same meteorite; (d) a spread of 1-3
Myr in radiometric ages of different meteorite types; and (e) a dearth of
melted asteroids, with model results for even some melted asteroids which imply
Myr delays in formation relative to ancient minerals. In recent years,
meteoritic evidence has appeared for some very early-formed planetesimals,
which represent a minority of both meteorites and asteroids. This implies that primary accretion started early and continued for several
million years - thus, it was fairly inefficient and did not run quickly to completion.  

Most current models for this ``primary accretion" stage (reviewed by Cuzzi and Weidenschilling 2006 and Dominik et al 2007) can be classified as either (a) incremental growth, where large particles sweep up smaller ones by inelastic collisions involving porous surfaces, and growth proceeds hierarchically; or (b) instability, where physical sticking is irrelevant and collective effects drive
collapse to km-sizes or larger on very short timescales. Those who favor
instability models, most of which rely on gravity and occur in a particle-rich
nebula midplane, are concerned by the poorly understood sticking of mineral
particle aggregates and the apparent difficulty of growing beyond meter size
due to rapid inward migration and collisional disruption. Those who
favor incremental growth have noted that midplane instability models are
precluded by even very weak global turbulence, and that, in the dense midplane
layers that form when turbulence is absent, incremental accretion is at low relative velocity and the meter-size
barrier is not a problem. However, the most sophisticated models of incremental
accretion in nonturbulent nebulae find it to be {\it so} efficient that large
planetesimals grow in only $10^4 - 10^5$ years throughout the asteroid belt
region (Weidenschilling 2000), a short time which is difficult to reconcile with constraints
(a-e) above. A third class of scenario suggests that a complex interplay between several nonlinear processes - turbulence, pressure gradients, and gravity - may concentrate appropriately sized particles and lead to planetesimal growth (Cuzzi et al 2001, 2005, 2007; Johansen et al 2006, 2007). Scenarios which involve preferentially meter-sized objects encounter concerns about whether meter-sized objects can survive their own high-velocity mutual collisions in this sort of turbulent environment ({\it cf.} Sirono 2000, Langkowski et al 2007, Ormel et al 2007). However, in principle, such boulder-concentration scenarios can proceed independently in the same environment as discussed in this paper, where we focus on mm-size particles. 

Most of the above scenarios provide no natural explanation for observed meteorite properties (a) and (b) above. In particular, the evidence for the H chondrite class (and perhaps all the ordinary chondrites) suggests that entire asteroids of 10-100km diameter formed directly from a physically, chemically, and isotopically homogeneous mix of dust-rimmed particles of similar size (section 2.1). In this paper, we outline a possible path by which entire batches of mm-size, aerodynamically sorted particles might proceed directly in turbulence (even if sporadically) to planetesimals having the properties outlined above. We find that sufficiently large and dense clumps of mm-size particles can form by turbulent concentration such that, even if classical gravitational instability can't operate, their self-gravity may still allow them to survive disruptive forces and slowly sediment into a ``sandpile" planetesimal. This simple analytical model is backed up by some numerical simulations that support the basic idea. In sections 2.2 and 2.3, we review the most relevant physics that determines the properties of dense, particle-rich zones or clumps in turbulent nebulae. In section 3, we address the fate of these dense clumps using analytical and numerical models of their evolution. We derive the combination of size and mass density a clump must have to evolve into a ``sandpile" having some degree of internal strength. In subsequent stages not modeled here, we imagine that collisions and thermal sintering transform these sandpiles into the cohesive rocky parent bodies we see today. However, in the outer solar system, lower energy collisions and weaker thermal processing might well allow planetesimals to retain their initial low-strength states, as seen for some cometary objects ({\it eg.,} Asphaug and Benz 1996). 

\section{Background}
\subsection{Meteoritics background and evidence for inefficient accretion in
a turbulent environment} One can extract a number of clues from primitive
(unmelted) meteorites regarding the primary accretion process by which their
parent bodies first formed (Scott and Krot 2005, Taylor 2005). 
For the best evidence, one must look back through
extensive subsequent evolutionary stages. Even unmelted bodies in the 100km
size range have incurred extensive collisional evolution (Bischoff et al.
2006), producing compaction, fragmentation, and physical grinding and mixing on
and beneath their surfaces, which may obscure the record of primary accretion.
Model studies ({\it e.g.}, Petit et al 2001, Kenyon and Bromley 2004, 2006;
Bottke et al 2005; Chambers 2004,2006; Weidenschilling and Cuzzi 2006) suggest
that the collisional stage occurred after dispersal of the nebula gas allowed
the orbital eccentricities of primitive bodies to grow. We are concerned with
an earlier stage, when the still-abundant nebula gas led to a more benign
environment with fewer and gentler collisions. The direct products of primary
accretion might be most clearly visible in the rare ``primary texture" seen in
some CM (Metzler et al 1992) and CO (Brearley 1993) chondrites. In these
objects, or more specifically in unbroken fragments within them, the texture
consists of similarly-sized, dust-rimmed particles packed next to each other as if gently brought together and compressed, with no evidence for local fracturing or grinding.

Even after collisional effects associated with subsequent stages of 
growth have blurred this signal, perhaps even mixing material from different
parent bodies, evidence remains in the bulk properties of all chondrite
classes. Chondrite classes are
defined by their distinctive mineral, chemical, and isotopic properties
({\it e.g.} Grossman et al 1988, Scott and Krot 2005, 
Weisberg et al 2006). Large samples of
material with a quite well defined nature were accumulated at one place and/or time, and material of a quite different, but equally well-defined, nature was accumulated at another place and/or time into a different parent body. Amongst
the most obvious aspects of these class properties is a dominance within
chondrites of sub-mm size mineral particles (generically ``chondrules";
Grossman 1988, Jones et al 2000, 2005; Connolly et al 2006) which are
aerodynamically well-sorted. Aerodynamic sorting of chondrite components 
was first emphasized by Dodd (1976), Hughes (1978, 1980), and Skinner and Leenhouts (1993), and has since been discussed by a variety of authors 
(see Cuzzi and Weidenschilling 2006 for a summary
of the observations and arguments supporting aerodynamics). Chondrules have a size which varies from class to class, but is distinctive and narrowly defined within a given class or chondrite. 

The evidence suggests that chondrules do not comprise merely a thin
surface layer swept up by a large object (Scott 2006). In the case which we suggest as an archetype, the H-type ordinary
chondrite parent body is widely believed to be a 100 km radius object (perhaps
the asteroid Hebe), initially composed entirely of a physically, chemically,
and isotopically homogeneous mix of chondrules and associated material, which
was thermally metamorphosed by accreted $^{26}$Al to a degree which varied with
depth, into an onion-shell structure, and subsequently broken up in several
stages (Taylor et al 1987, Trieloff et al 2003, McSween et al 2002, Grimm et al
2005). Bottke et al (2005) conclude from models of collisional evolution in the primordial (massive) and the current (depleted) asteroid belt, that the primordial asteroid mass distribution was dominated by objects having diameter of around 100km, rather than having a powerlaw size distribution somewhat like that of the current population. 

We see the essential challenge as understanding how such a large object can be
assembled from such a homogeneous mixture of mm-size constituents, while other objects
(arguably the parents of the L and LL type ordinary chondrites as well, and logically then the parents of the 
chondrites of all classes) are assembled from distinct, yet comparably
homogeneous, ensembles of qualitatively similar particles. 
Moreover this assembly, or primary
accretion, phase of nebula evolution must persist for a duration of 1Myr or more between the formation times of the oldest and youngest meteorites (Wadhwa et al 2007) and even of objects found in the same meteorite. The $>$ 1 Myr age spread between ancient refractory inclusions (CAIs) and chondrules is well known (Russell et al 2006), but there may even be a comparably extended age spread between different chondrules in a given meteorite (Kita et al 2005).   

In recent years, radiometric dating of achondrites (melted meteorites) has
revealed some early-formed (and early melted) planetesimals (Kleine et al 2005,
2006). One expects early formation and melting to go together, because objects
larger than only 10km or so, which accreted at the same time that refractory
inclusions formed (4.567 Byr ago), would have incorporated their full
complement of $^{26}$Al and would have melted (LaTourette and Wasserburg 1998;
Woolum and Cassen 1999, Hevey and Sanders 2006). Thus, early accretion implies melting, and lack of
melting requires late accretion. Model results imply that, to avoid complete
melting, 100km bodies must accrete only after a typically 1.5-2 Myr delay
relative to ancient CAI minerals (reviewed by McSween et al 2002 and Ghosh et
al 2006). While unknown selection and sampling effects may influence the limited meteorite data record, spectral analysis of the entire asteroid database also suggests that melted asteroids are rare. That is, while debate continues as to whether small amounts of melt might be found on the surfaces of many asteroids, which might be caused by impacts ({\it e.g.} Gaffey et al 1993,
2002), only Vesta and a handful of other examples of fully melted basaltic
objects have been found in spite of extensive searches (Binzel et al 2002;
Moskovitz et al 2007).

Put together, this body of information suggests that primary accretion started
early but continued for several million years - thus, it was inefficient, 
or at least sporadic. By 
comparison, midplane incremental accretion in a nonturbulent nebula accretes numerous Ceres size bodies and tens of thousands of 10-100km size objects in only $10^5$ years (Weidenschilling 2000); thus, it is highly efficient. 

Models suggest that
the density, temperature, and composition of the nebula evolved significantly
over several million years (Bell et al 1995, D'Alessio et al. 2005, Ciesla and
Cuzzi 2006). Thus one expects the properties of forming planetesimals to change
with {\it time} even though the chemical, isotopic, and mineralogical
properties of the mixture of solids from which they formed might have been
fairly uniform in {\it space} at any {\it given} time (see Cuzzi et al 2003 and Ciesla 2007  for more detailed discussions). Reality might have been even more complicated - there is some evidence that chondrites of very distinct chemical and isotopic properties might have formed at roughly the same time (Kita et al 2005), suggesting a combination of spatial {\it and} temporal gradients.

In this paper we present the overview of a scenario by which accumulation of a well-defined particle mix into a large planetesimal - making it a ``snapshot" of the local nebula mix - might occur sporadically or inefficiently in weak turbulence, over an extended period of time and throughout the terrestrial planet region. The challenge for the future is to show that the accretion processes discussed here can {\it quantitatively} account for the mass needed to build planetesimals of the mass needed to create planetary embryos and planets during the age of the nebula (Chambers 2004, 2006; Cuzzi et al 2007).

\subsection{Turbulence} A variety of astronomical and meteoritical evidence
supports weak, but widespread and sustained, nebula turbulence. The observed abundance of
small particles at high altitudes above the midplane in many visible
protoplanetary nebulae for millions of years is most easily explained by
ongoing turbulence - both to regenerate the small grains in collisions and to
redistribute them to the observed altitudes (Dullemond and Dominik 2005). 
The survival of ancient,
high-temperature mineral inclusions (CAIs) in chondrites, in spite of their
tendency to drift inwards towards the sun, can also be explained by turbulent
diffusion (Cuzzi et al 2003, 2005). Moreover, the recent discovery by STARDUST of both high- and
moderate-temperature crystalline silicates in comet Wild2 (Brownlee et al 2007, Zolensky et al 2007) can be easily understood in this way (Ciesla 2007).

It remains a subject of active theoretical debate as to how, or even whether, the nebula can maintain itself in a turbulent state. Ongoing dissipation by molecular
viscosity $\nu_m$ on the smallest, or Kolmogorov, lengthscale requires a
production mechanism to continually generate turbulent kinetic energy. Although the very existence of ongoing disk accretion into the star, with the ensuing
gravitational energy release, provides an ample energy source, the vehicle
for transferring that energy into turbulent gas motions is undetermined. Baroclinic instabilities require substantial opacity at thermal wavelengths, but might be suitable to generate turbulence during the fairly opaque chondrule-CAI epoch in the inner solar system (Klahr and Bodenheimer 2003). The widely accepted magnetorotational instability (MRI) (Stone et al 2000) may or may not be precluded at the high gas densities of the inner solar system (Turner et al 2007). The role of pure hydrodynamics operating on radial shear remains unclear, because Keplerian disks are nominally stable to linear (small) perturbations (Balbus et al 1996, Stone et al 2000). However, a number of current theoretical studies suggest turbulence might indeed be present ({\it eg.} Arldt and Urpin 2004, Busse 2004, Umurhan and Regev 2004, Dubrulle et al 2005, Afshordi et al 2005, Mukhopadhyay et al 2005, Mukhopadhyay et al 2006). Much of this recent work focusses on nonlinear (finite amplitude) instabilities occurring at the very high Reynolds numbers characterizing the nebula, which are well beyond the capability of current numerical and even laboratory models. 

%\clearpage
%\begin{figure*}[t].. below for single column width
\begin{figure}[t]                              
\centering                                     
\includegraphics[angle=0,width=3.1in,height=3.1in]{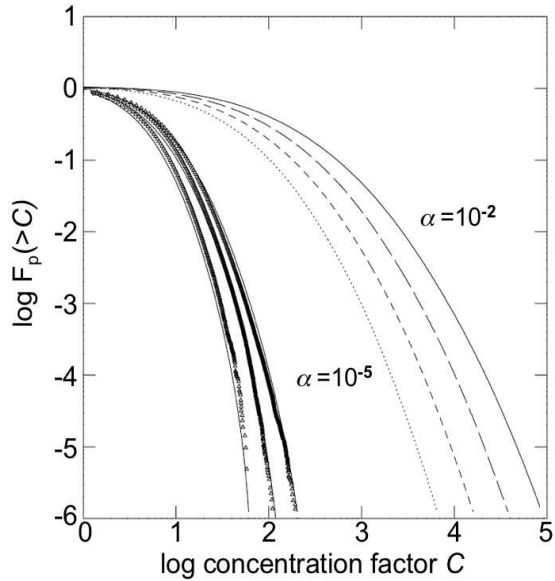}
%\vspace{-0.2in}
\caption{Cumulative probability distribution, or volume fraction $F_p$, of particles
having local concentration factor larger than some value of $C$ (from Cuzzi et al 2001; see also section 2.3).  
The curves with symbols
represent comparison of the predictive theory described in Cuzzi et al (2001) with 3D numerical
simulations at moderate Reynolds numbers; the curves without symbols are extension of the same theory to
nebula conditions denoted by their values of $\alpha$. Concentration factors $C \sim 100$ of chondrule precursors
(leading to $\Phi = \rho_p/\rho_g \sim 1$) are fairly common for a range of
nebula turbulent intensities (denoted by $\alpha$), and help explain lack of
chondrule isotopic fractionation (Cuzzi and Alexander 2006). In this paper we focus 
on the much less common, but much denser, regions towards the bottom right side of the plot
($C\sim 10^4$ or $\Phi \sim 100$). More general PDFs which separate out the role of vorticity and incorporate the effect of particle mass loading are discussed in section 2.3 and Appendix A. }
\end{figure}
%\end{figure*}
%\clearpage

In this paper, we simply assume the presence of weak nebula
turbulence. The intensity of turbulence may be characterized by a
nondimensional parameter $\alpha = (v_t/c)^2$, where $v_t$ is the
velocity of the most energetic eddies and $c$ is the sound speed. Values of
$\alpha$ in the range $10^{-4 \pm 1}$ are consistent with observed nebula
lifetimes, and can be sustained by only a few percent of the gravitational
energy released as the nebula disk flows into the sun. One then estimates the
most energetic (largest) eddy size as $L \sim H \alpha^{1/2}$, where $H$ is the
nebula vertical scale height, or disk thickness. A cascade to smaller scales
ensues, where the smallest (Kolmogorov) eddy scale is $\eta = L {\rm Re}^{-3/4} \sim$
1 km and ${\rm Re} = \alpha c H /\nu_m \sim 10^7$ for $\alpha\sim 10^{-4}$ is the
nebula Reynolds number (Cuzzi et al 2001). 

\subsection{Turbulent concentration} It has been found both numerically and
experimentally that particles of a well defined size and density are
concentrated into very dense zones in realistic 3D turbulence (Eaton and Fessler 1994). 
The
optimally concentrated particle has a gas drag stopping time $t_s$ which is
equal to the overturn time of the Kolmogorov scale eddies. In the nebula, $t_s
= r \rho_s/c \rho_g$ where $r$ and $\rho_s$ are the particle radius and
material density, and $c$ and $\rho_g$ are the gas sound speed and density.  The
Kolmogorov eddy frequency $\omega_{\eta}$ scales with the orbital frequency
$\Omega$ as $\omega_\eta = \Omega {\rm Re}^{-1/2}$, and particles are most
effectively concentrated when $t_s \omega_\eta \sim 1$ (Eaton and Fessler 1994). 
Chondrule-size silicate particles are concentrated for canonical nebula properties if $\alpha \sim 10^{-5}-10^{-3}$; moreover the {\it shape} of the concentrated particle size distribution is parameter-independent, and agrees very well with that of chondrules (Paque and Cuzzi 1997, Cuzzi et al 2001). We define the concentration factor $C \equiv \rho_p/\overline{\rho_p}$, where we denote the local volume mass density of particles as $\rho_p$ and $\overline{\rho_p}$ is its
nebula average value, and the particle mass loading $\Phi \equiv \rho_p/\rho_g = C \overline{\rho_p}/\rho_g$. For instance, figure 1 (Cuzzi et al 2001) shows the probability distribution function (PDF) for $C$ determined from numerical simulations at moderate Reynolds number (curves with symbols) and predicted for plausible nebula Reynolds numbers (curves without symbols). Note that the volume fraction having $C > 100$ is quite substantial, which helps explain some mineralogical and isotopic properties of chondrules (Cuzzi and Alexander 2006).{\footnote {Here, $\Phi=\rho_p/\rho_g$ is different from the definition in Cuzzi and Alexander (2006), which is equivalent to $\rho_p/A \rho_g$, where $A \sim 10^{-2}$ is the  fractional abundance of solids to hydrogen gas by mass. Thus values of $\Phi$  in Cuzzi and Alexander (2006) are the equivalent of $\rho_p/\rho_g \sim 1$ - a much more common occurrence than the situation we study here.}} Not surprisingly, the volume fraction decreases for larger $C$. In this work we focus on the less abundant, but higher mass density, concentrations towards the lower-right hand side of the PDF. 

The turbulent concentration PDFs of figure 1 did not allow for the feedback of the concentrated particles on the gas. This ``mass loading" would be expected to damp gas turbulent motions once the particle mass density significantly exceeds that of the gas, and lead to a saturation of concentration at some value. Recent work which {\it includes} the feedback effects, through drag, of particle density on the damping of turbulence, has shown that values of $\Phi \leq 100$ can indeed be achieved - although less commonly than in the unloaded models such as shown in figure 1.  This {\it cascade model} of turbulent concentration (Hogan and Cuzzi 2007) is summarized in Appendix A. Based on these results, we adopt $\Phi = 100$ as an upper bound in all stability and evolution modeling. 

\section{Primary accretion of planetesimals}
The ability of gravity to overwhelm opposing forces is a well-known theme in astrophysics, dating back 80 years to early studies of star formation by Jeans. Over 30 years ago, gravitational instability (GI) of solids in a dense
layer near the nebula midplane was proposed to lead directly to solid
planetesimals on a dynamical timescale (the orbit time) (Safronov 1969, 
Goldreich and Ward 1973). 
Traditional GI thresholds require
the dynamical (collapse) time of a dense region $t_G = \pi (G \rho_p)^{-1/2}$,
where $G$ is the gravitational constant, to be shorter than both the transit
time due to random velocity and the local shear (vorticity) timescale (Toomre 1964).
These arguments imply that GI occurs when the local particle density exceeds a
few times the Roche density $\rho_R = 3 M_{\odot}/4 \pi a^3$, where $M_{\odot}$ 
is the solar mass and $a$ is the distance from the sun (Safronov 1991, Cuzzi et al 2001). However, this
scenario is frustrated by self-generated midplane turbulence even if the nebula
is not globally turbulent ({\it e.g.} Weidenschilling 1980,1995; Cuzzi et al 1993,1994; Johansen et al 2007). That is, a dense layer of cm-size and larger particles orbits at a different velocity than the pressure-supported gas, leading to a vertical velocity shear which becomes turbulent and stirs the particle layer. 
A more recent flavor of GI requires
the particles to be small enough that the particle-rich midplane acts as a
single fluid, like fog, so that its vertical density stratification stabilizes
it against self-generated turbulence (Sekiya 1998, Sekiya and Ishitsu 2001, Youdin and Shu 2002, Sekiya and Takeda 2003, Youdin and Chiang 2004). Because the particles must be very small to satisfy this {\it one-phase} requirement (mm-size and smaller), their settling towards the midplane is frustrated by even the faintest breath of nebula turbulence ($\alpha \sim 10^{-9}$). If the particles are prevented from settling into a sufficiently dense layer that their mass density can stratify the total density (eg, Dubrulle et al 1995, Cuzzi and Weidenschilling 2006), the conditions needed to trigger the instability are never achieved. 

In the following subsections, we study the evolution of a dense clump such as might form towards the lower right domain of figure 1, albeit in a simplified way. We treat a clump as an isolated object when in reality it exists in a context of surrounding material of varying density. In section 4, we discuss the implications of more realistic conditions. 

\subsection{Gas pressure precludes particle gravitational instability}
The role of the {\it gas pressure} in GI for nebula particles has been almost universally ignored, even though its importance was pointed out decades ago by Sekiya (1983). We encountered it ourselves independently. 
To test our GI thresholds, we initially ran numerical models of static, but very dense,
clumps ($\Phi = 1000$), expecting them to collapse with
gravitational free-fall or dynamical collapse times $t_G = \pi (G \rho_p)^{-1/2}
= \pi (G \Phi \rho_g)^{-1/2}$ and velocities $V_G \approx l (G
\rho_p)^{1/2}/2\pi$, where $l$ is the clump diameter. We assumed a spherically symmetrical dense particle clump with density $\rho_p$,
embedded in gas with density $\rho_g$, where $\Phi = \rho_p/\rho_g$. Instead, only very slow shrinkage ensued.  We suspected that the collapse was being artificially blocked by our incompressible code. 

We then developed a fully compressible 1D model which
confirmed our incompressible calculations and clearly demonstrated a much higher, gas-pressure-dependent $\Phi$ threshold for GI in the limit when $t_s \ll t_G$.
Results from the model are shown in figure 2. 

%\clearpage
%\begin{figure*}[t].. below for single column width
\begin{figure}[t]                                 
\centering                                                                   
\includegraphics[angle=0,width=3.1in,height=2.5in]{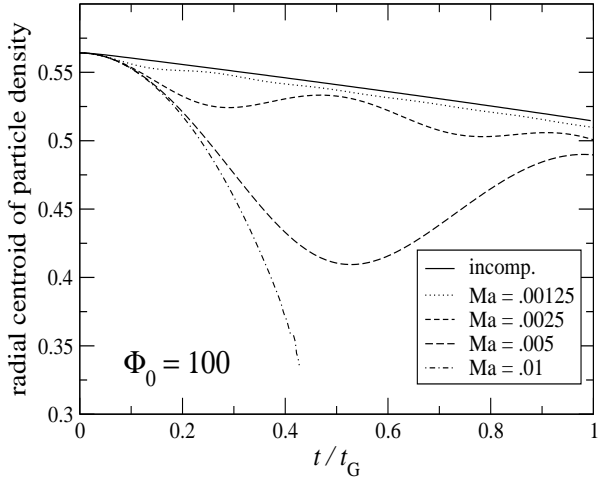}
\vspace{0.2 in}
\caption{Results of a 1D compressible numerical model, showing how gas pressure
precludes dynamical timescale gravitational instability (Sekiya 1983) as described
in more detail in section 3.1.  Here, ${\rm Ma} = (l/c)(G \rho_g)^{1/2} =
1/\Phi^*$ is like a Mach number.  Nebula values of $c$ and
$\rho_g$ lie in the top set of nearly linear curves,
allowing only slow inward sedimentation on the timescale $t_{sed}$ 
(for $\Phi_0 = 100$). That is, under nebula conditions of gas density and temperature, and particle loading, the gas behaves as if it were incompressible and prevents dynamical timescale GI for particles with stopping time much less than the dynamical collapse time.}
\end{figure}
%\end{figure*}
%\clearpage

{\bf Figure 2} shows results
from this fully compressible model, illustrating how $\Phi > c/(l (G
\rho_g)^{1/2})$ is required for traditional (dynamical timescale) instability to occur. The initial condition was a Gaussian blob having density distribution 
$\rho_p(r) = \Phi_0 \rho_g \exp(-r^2/l^2)$.  Slow shrinkage of the particles through the gas is seen in the stable, ``incompressible" regime where nebula parameters lie, well within the flat curves at the top of the plot with  ${\rm Ma} \sim 10^{-4}$ for $l\sim 10^4$km, $c \sim 10^5$cm/s, and $\rho_g \sim 10^{-9}$g cm$^{-3}$. The high value of ${\rm Ma}$ required for true dynamical timescale collapse (given the chosen value of $\Phi = 100$) would require smaller gas sound speed or larger clump size $l$ and gas density $\rho_g$, by orders of magnitude. 

The slow shrinkage in the stable regime results from particles sedimenting inwards towards their mutual center under their own self-gravity at their terminal velocity $v_T = g t_s$, where
$g = 2 G \Phi l \rho_g$ is the local gravitational acceleration due to the clump's mass, and $t_s =
r \rho_s/c \rho_g$ is the particle stopping time. Thus $v_T = 2 G \Phi l r \rho_s/c  \ll V_G$ for $t_s \ll t_G$, and the ``sedimentation" timescale is 
\begin{equation}
t_{sed} = {l \over 2 v_T} ={ 1 \over 4 G \Phi (r \rho_s/c)} = { c \over 4 G \Phi  r \rho_s } = { 1 \over 4 G \Phi \rho_g t_s},
\end{equation}
roughly 30-300 orbit periods at 2.5 AU for 300$\mu$ radius chondrules
and $\Phi = 1000 - 100$; that is, much longer than $t_G$. It seems inappropriate to
describe this slow ongoing evolution as an instability - certainly in midplane scenarios where it was slow vertical sedimentation by individual particles toward the midplane that produced the ``unstable" situation in the first place, and where only further slow sedimentation transpires. The situation is more akin to star formation mediated by ambipolar diffusion than by traditional gravitational instability. Gas pressure even precludes the
Safronov-Goldreich-Ward GI in the form it was originally proposed, where
cm-sized particles were envisioned (although $t_{sed}$ would be faster for 
literally cm-sized particles, and more closely approaches $t_G$).

The results of Sekiya (1983) have apparently been overlooked by all subsequent workers, who have invariably assumed a Toomre-Safronov-Goldreich-Ward type {\it particle} ensemble which is decoupled from the gas, and stabilized (on small scales) only by particle random velocities. This gas pressure constraint is, however, crucial for particles in the chondrule size regime where $\ts \sim$ an hour and
$t_G \sim$ a year, and must be applied to all GI models involving
chondrule-sized particles. The good news is that for the interesting range of
nebula gas density and particle concentration values, the gas {\it does}
actually behave as if it were incompressible (validating the use of an incompressible code in numerical models such as shown in section 3.4).

The physics is easily understood in the one-phase regime $(t_s \ll t_G)$. The
particles feel no direct pressure force, but are firmly trapped to the gas
which does. The (inward) gravitational force is dominated by particles for
$\Phi \gg 1$. The particles drag and compress the gas with them as they begin to
collapse under their self-gravity, producing a radial gas density gradient $d
\rho_g /d l$; this translates into a large outward {\it pressure} gradient $c^2
d \rho_g /d l$ which acts on the gas, and thereby also on the strongly coupled
particles.

The gravitational force per unit volume is $f_G = 4 G M \rho_p /l^2 \sim 2G \Phi^2 \rho_g^2 l$, where $M$ is the clump mass and $R$ its radius. After some
incipient shrinkage and compression occurs, the outward pressure gradient force
per unit volume becomes roughly $f_P = c^2 d\rho_g/dl \sim c^2 \rho_g/l$, where $c$
is the gas sound speed. Then for GI to be possible, $f_G/f_p = G \Phi^2 l^2
\rho_g /c^2 > 1$, giving the criterion for GI as $\Phi > \Phi^*$, where $\Phi^* =
c/l(G \rho_g)^{1/2}$. $\Phi^*$ is like an inverse Mach number ${\rm Ma}^{-1}$,
where ${\rm Ma}  = (l/c)(G \rho_g)^{1/2}$. An
alternate (perturbation) approach compares the incremental changes $\delta f_p$
and $\delta f_G$ associated with a small gravitationally induced shrinkage
characterized by $\delta l \ll l$. It is straightforward to show (for $t_s \ll
t_G$) that $\delta f_p/\delta f_G \approx c^2 \rho_g / G M \Phi \sim 10^5$
under nebula conditions. Thus the clump is
``stiffened" by pressure and even a small incremental shrinkage is
self-limiting. Inspection of {\bf figure 2} shows small
oscillations which suggest that a formal linear stability analysis might
lead to an improved stability criterion; this is, however, beyond the scope of the present paper.

For $l \sim$ a few $10^3$ km and $\rho_g \sim 10^{-9}$ g cm$^{-3}$ (Desch et al
2005), $\Phi^* \sim 10^4$. This criterion for GI can be compared with the
traditional criterion for (marginal, or maximal) instability which is widely
used as {\it the} criterion for gravitational instability, roughly twice the Roche
density $\rho_R = 3 M_{\odot}/4 \pi a^3$ (Goldreich and Ward 1973, Safronov
1991, Cuzzi et al 1993). The ratio of the two thresholds is:
$$ 
{ \Phi^* \rho_g \over 2 \rho_R} = {\rho_g \over 2 \rho_R { \rm Ma}} \approx 
{ 2c (G \rho_g)^{1/2} \over \Omega^2 l} \sim { 2-6 \times 10^3 \over (l/{\rm 10^3
km})},
$$ 
for $\rho_g =10^{-10}$ to $10^{-9}$. For the midplane GIs which have been widely discussed in the past (reviewed by Cuzzi and Weidenschilling 2006), $l \sim 10^2$km, so the traditional GI criterion falls short by a factor of more than $10^4$ ({\it cf.} Sekiya 1983)! This degree of mass concentration is unlikely under any plausible
circumstances, especially for particles which are already well trapped to the
gas. For the ubiquitous dense clumps we discuss here, which form at all elevations, $l \sim 10^4$ km. Thus the pressure-supported gas phase prevents the tightly coupled particles from undergoing GI until the particle mass loading is hundreds of times larger than the traditional GI criterion (which is on the order of the Roche density
$\rho_R$).  

Nevertheless, Sekiya (1983) also noted that when $\rho_p$ is in the range usually
cited for GI (a few times $\rho_R$, or $\Phi \sim$ a few hundred), a 3D
``incompressible mode" arises which, while not explicitly stated, we interpret as retaining 
an identifiable cohort of particles. An ``incompressible mode" might resemble a blob of water oscillating in zero-gravity. Dense zones which
result from turbulent concentration are candidates for such incompressible
modes, even at mass loadings too low to induce actual GI. However, the fate of such slowly evolving entities in the presence of likely perturbations has never been explored.
Below we discuss the least avoidable, most pervasive disruptive perturbation which such clumps would encounter once they form, which we believe to be ram pressure disruption due to systematic velocity differences between the dense zones and the gas, and then derive conditions under which these clumps can survive to become planetesimals.

\subsection{The fate of ``incompressible" clumps}

Several kinds of perturbation by the
enveloping nebula gas might disrupt a dense clump  before the slow sedimentation
of its constituent particles towards their mutual center (on the timescale $t_{sed}$) can produce a
moderately compact object. 

Perhaps the simplest to discuss and dismiss are turbulent pressure, velocity, and/or vorticity {\it fluctuations} encountered by the clump. Turbulent pressure  fluctuations on the scale $l$ have typical intensity $\rho_g v_t(l)^2$,
where $v_t(l)$ is certainly smaller than the fluctuating velocity $v_t(L) = c 
\alpha^{1/2}$ of the largest eddy (the dominant energy-containing eddy). The
ratio of even the largest eddy pressure fluctuations to the steady ram pressure from the headwind $\beta \Omega a = \beta v_K$ ($a$ is the semimajor axis) 
is $(v_t/\beta v_K)^2 = \alpha c^2
/\beta^2 v_K^2$. Since $c/v_K = H/a$ and $\beta = (H/a)^2$, $(v_t/\beta v_K)^2
< \alpha/\beta < 1$ unless $\alpha > 10^{-3}$. Nebula evolutionary
timescales, and parameters leading to turbulent concentration of
chondrule-sized particles in the asteroid belt region, imply $10^{-6} < \alpha < 10^{-3}$ depending on gas density and other properties (Cuzzi et al 2001).
This suggests that turbulent velocity fluctuations on clump lengthscales play a small role. We have run simulations of clumps settling in gravity with and without enveloping turbulence, and the typical evolutions are qualitatively similar. 

We next address local {\it average} vorticities on the lengthscale of a clump. It has long been known that local vorticity is a factor in gravitational instability ( Toomre 1964, Goldreich and Ward 1973). Because eddies much smaller than the
largest scale of turbulence typically have larger vorticity than the Keplerian
shear commonly explored ({\it eg.} Toomre 1964), this is a concern in principle. For example, in the inertial range, 
$\omega(l) \sim \Omega(L/l)^{2/3}$ (Tennekes and Lumley 1972). Then for 
$l \sim 10^4$km, $L \sim H \alpha^{1/2}$ and $\alpha \sim 10^{-4}$, $\omega(l) \sim 10^4 \Omega$. However,
turbulent concentration has the property that dense particle zones
preferentially lie in zones of locally {\it low} vorticity compared to the
average at their lengthscale (Eaton and Fessler 1994). Statistical studies
(Hogan and Cuzzi 2007) show that
dense zones in high ${\rm Re}$ environments can form in regions with local
vorticity 1-2 orders of magnitude smaller than the average value expected for that lengthscale (see also figure 5). Below we sketch how such a constraint is derived and applied. 

A simplified requirement for gravitational binding of a clump of lengthscale $l$ and mass density $\rho_p$, in the presence of local rotation at angular velocity $\omega(l)$, is $4 G \rho_p > \omega^2(l)$. Recalling that the local vorticity $\omega(l)$ can vary by orders of magnitude from its average value $\langle \omega(l)\rangle$, we normalize both sides of the above relationship between $\rho_p$ and $\omega$ by the average inertial range enstrophy $\langle \omega^2(l)\rangle$, and define the normalized enstrophy $S = (\omega^2(l)/\langle \omega^2(l)\rangle)$. This quantity provides the horizontal axis on figure 5. Then the simple expression above can be rewritten as $\Phi > \langle \omega^2(l)\rangle S / 4 G \rho_g$. Inertial range relationships can be used to express $\langle \omega^2(l)\rangle = \Omega^2(L/l)^{4/3} = \Omega^2(L/B \eta)^{4/3}$ where $\eta$ is the Kolmogorov scale and $B$ is some scaling factor. Making use of 
${\rm Re} = \alpha c H /\nu_m = (L/\eta)^{4/3}$ as discussed earlier, and substituting nebula parameters at 2.5 AU, we find that gravitational binding requires 
$\Phi > 3 \times 10^6 \alpha S$. This constraint yields diagonal lines on plots of the sort shown in figure 5 (for an example see Cuzzi et al 2007). Interesting regions of parameter space remain accessible to a degree that depends on quantitative modeling of the PDF $P(\Phi,S)$. 

Obviously, more detailed numerical models of clumps in realistic turbulence are needed to assess these perturbations. We will assume, for the purposes of this study, that turbulent pressure (and associated vorticity) fluctuations are a minor
influence and that regions which are stable to their own local vorticity exist (see Cuzzi et al 2007). This leaves the dominant disruptive process, unavoidably shared by turbulent and laminar nebulae, as ram pressure disruption due to {\it systematic} velocity differences between the clump and the gas, as described below. Incidentally, survival against ram pressure gives the horizontal stability boundaries in figure 3 of Cuzzi et al (2007). Emplacing these constraints on PDFs obtained from cascade models is somewhat involved, and a full quantitative description of the situation is deferred to a future paper.

\subsection{Ram pressure disruption and the Weber number}
Here, for simplicity, we envision the evolution of a single, isolated dense clump with $\Phi \gg 1$, merely as an example of one of the many dense clumps formed near the lower right hand corner of figure 1. The clump exerts a strong collective influence on the entrained gas, and the mixture experiences solar gravity as a unit. It attempts
to move as an individual large object relative to the nebula gas, dragging the
entrained gas along. For instance, a clump forming at some distance above the
nebula midplane settles downward under the acceleration of the vertical
component of solar gravity, whereas the surrounding gas remains 
under vertical hydrostatic balance. Also, nebula gas generally orbits at a
slower velocity than the Keplerian orbit velocity $v_K$ obeyed by massive
particles. This is because of its generally outward radial pressure gradient
force $-(1/\rho_g)dP/da = 2\beta \Omega a^2$, where $a$ is the distance from
the sun, and $\beta \sim 10^{-3}$). Thus, particles experience a headwind with
magnitude $w_g \sim \beta \Omega a = \beta v_K \sim$ a few $10^3$ cm/sec. These
headwinds result in a ram pressure $\rho_g w_g^2/2$ which can disrupt a
strengthless clump.

For example, a dense drop of ink settling in a glass of water is quickly
disrupted by the Rayleigh-Taylor instability and mixes with its
surroundings (Thomson and Newhall 1885). However, a dense drop of fluid {\it with surface tension} can
avoid disruption and settle indefinitely at terminal velocity (as do raindrops,
or water in oil). In this situation, the criterion for stability of a fluid
droplet, which determines its terminal velocity and thence its size, is given
by the Weber number ${\rm We}$, where $ {\rm We} = 2 r \rho_g v^2 / \gamma$,
and $\gamma$ is the surface tension coefficient. Drops are disrupted when ${\rm
We}$ exceeds some critical value ${\rm We}^* \sim 1-10$ (Pruppacher and Klett 1997).

Dense particle clumps in the nebula have no surface tension, but they do have
self-gravity. By direct analogy with the droplet surface tension criterion, we define a {\it gravitational Weber number} ${\rm We}_G$ as the ratio of the ram pressure force per unit area to the self-gravitational force per unit area for a flattened disk, which is the initial stage of a strengthless spherical clump of diameter $l$ and particle density $\rho_p = \Phi \rho_g$ upon encountering a headwind $w_g$ (Thomson and Newhall 1885). For a disk of surface mass density $\sigma_p = \rho_p l$, the gravitational force per unit area is $G \sigma_p^2$. Then 
\begin{equation}
{\rm We}_G = {C_D \rho_g w_g^{2} \over 2 G \sigma_p^2} = {C_D w_g^2 \over 2 G
\Phi^2 \rho_g l^2},
\end{equation}
where $C_D$ is an effective drag coefficient for the clump, on the order of unity (see Appendix B). $We_G$ can be written in other ways as well. The orbit frequency at $a$ provides the useful relationship $\Omega(a) = (G M_{\odot} /a^3)^{1/2} \equiv (G \rho^*)^{1/2}, \hspace{0.1 in} {\rm or} \hspace{0.1 in} G=\Omega^2/\rho^*$, where $\rho^* = M_{\odot}/a^3 \sim 3.8 \times 10^{-8}$ g cm$^{-3}$ at 2.5 AU\footnote{Note $\rho^*$ is different from Safronov's (1991) Roche density $\rho_R = 3 M_{\odot}/ 4 \pi a^3$ discussed earlier}; then
\begin{equation}
{\rm We}_G = {C_D  \rho^*  w_g^2  \over 2 \Phi^2 \rho_g \Omega^2 l^2}
= {C_D \beta^2 \over  2 \Phi^2}{ \rho^* \over \rho_g}{ a^2 \over l^2}.
\end{equation}
By analogy with the more familiar surface tension case, there will be some critical {\it gravitational} Weber number for stability ${\rm We_G}^*$, that is probably on the order of unity (Pruppacher and Klett 1997); its value must be constrained by numerical experiments as described below. Then we require for stability against headwind disruption that
${\rm We}_G < {\rm We}^*_G$ or 
\begin{equation}
\Phi l > w_g/(2 G {\rm We}_G^* \rho_g)^{1/2}
  = {\beta a \Omega \over (2 G \rho_g {\rm We}^*_G/C_D)^{1/2} }.
\end{equation}
These expressions determine the combination of clump diameter $l$ and particle
loading $\Phi$ that will stabilize it against a headwind of magnitude $\beta
v_K$. The headwind due to nebula radial pressure gradient will occur even if a clump is at the midplane and has zero settling velocity; this gives the lower limit on ram pressure that the clump must be stable against. Neglecting vertical settling restricts potentially stable clumps to lying within about 0.01 gas scale heights of the midplane, where the headwind $\beta \Omega a$ is comparable to the vertical settling velocity. This restriction must be factored into statistical estimates of planetesimal  production (see, e.g. Cuzzi et al 2007). 

For typical nebula parameters, we assume gas density  to be in the range $\rho_g = 10^{-10}$g cm$^{-3}$ (a nominal minimum mass value) to $10^{-9}$ g cm$^{-3}$ (a more
recent value supported by nebula evolution and chondrule formation; {\it cf.}
Desch et al 2005), semimajor axis $a = 3.8 \times 10^{13}$ cm (2.5 AU),
pressure gradient/headwind parameter $\beta \sim 10^{-3}$ (see, {\it eg.}, Nakagawa et al 1986, Cuzzi et al 1993),
and solar mass $M_{\odot} = 2 \times 10^{33}$g. In this regime, $\Phi l > 1.5-5 \times10^6(\beta/10^{-3})$  km. Recall from section 2.3 and Appendix A that mass loading limits the maximum achievable mass loading to $\Phi \sim 100$. A clump satisfying the above constraint, with $l = 1-5 \times 10^4$ km, would have the mass of a 10-100 km radius body of unit density - thus, these precursors can lead to quite sizeable objects. This characteristic size range is intriguingly close to the roughly 50 km radius primordial building block size of Bottke et al (2005). The very existence of {\it any} preferred size for the primordial population, if true, is an intriguing result. 

\subsection{Numerical model of clump evolution} To obtain a sanity check on the concepts of section 3.3, and to constrain the (unknown) value of ${\rm We}_G^*$ for this problem, we developed
a very simplified numerical model of a clump experiencing a steady nebula headwind from a more
slowly orbiting, pressure-supported gas, using Hill's approximation which
transforms a cylindrical system into a cartesian system rotating at some mean
rate ${\bf \Omega_0}$ - useful if the domain covers a sufficiently narrow
radial and angular region that the curvature is negligible. The overall orbital
motion of particles (the Keplerian velocity) is thus subtracted out, and the
gas exhibits a small differential velocity because of its deviation from
Keplerian. A particle clump under influence of the gas will slowly lag behind in the $y$ (orbital)
direction, and slowly drift to smaller $x$ (radial) values, due to the
gas headwind. The model is not intended to be a high-fidelity representation of a realistic nebula situation, but merely to check the basic Weber number model of section 3.3.
 
The code evolves the velocity field of a two-phase system. The gas is perturbed by a clump of particles, which are mutually attracted by gravitational forces and are themselves dragged by, and exchange momentum with, the gas. The Hill frame,
with cartesian coordinates $(x,y,z)$, represents a radially narrow region
corresponding to some range of radius $a$, orbital angle $\theta$, and vertical
distance $Z$, respectively. Periodic boundary conditions are used in all three dimensions. The instantaneous Navier-Stokes (Eulerian) equations describing the
conservation of mass and momentum for the incompressible gas are expressed in
the Hill frame as,
 
\begin{equation}
  \mathbf{\nabla \cdot U} = 0,
\end{equation}
and

\begin{eqnarray}
  \frac{\partial{\mathbf{U}}}{\partial{t}} + \mathbf{(U \cdot \nabla ) U} 
   & = &\- \frac{\mathbf{\nabla} P + \mathbf{\nabla}P^{glob}}{\rho_g} + \nu
\mathbf{\nabla^2 U} \nonumber \\
   &   &   - 2 \mathbf{ \Omega_0} \times {\mathbf {U}} \  
+ \mathbf{f}^{drag} + \mathbf{f}^{sp}
\end{eqnarray}
where $\mathbf{U}$ is fluid velocity, $\rho_g$  is gas mass density, $\nu$
is gas viscosity, $P$ is local pressure (which fluctuates along with gas
velocity and vorticity variations), $\mathbf{\nabla}P^{glob} = ( 2 \rho_g
w_g\Omega_0, 0,  0) $ represents the (constant) global nebula radial pressure
gradient, and $\mathbf{\Omega_0} = (0, 0,\Omega_0)$ is the Kepler frequency
vector. This
setup causes the gas to flow towards the clump in the $y$ direction at a uniform speed $w_g$. We are
currently neglecting the Keplerian radial shear term $2 q \Omega^2 x
\mathbf{e_x}$, where $\Omega(a) = \Omega_0(a/a_o)^{-q}$ and $\Omega_0=
\Omega(a_0)$, in the gas and particle equations to get a better comparison with
our analytical stability model. The shear rate associated with this term
$(dw/dx)_{Kep} = \Omega_0/2$, whereas a typical (fluctuating) local shear or
strain due only to turbulent motions is $(dw/dx)_{turb} \sim \omega(l) >
\Omega_0$. Because the turbulent shear is so widely varying, and likely to be larger than the Keplerian shear, in this study we neglect both of these complications,
though they are both suitable avenues for future work.
 
In our code, particles are followed in the Lagrangian sense. The term $\mathbf{f}^{drag}$ represents the force per unit mass imparted to
the gas by the particles. It has the general form
\begin{equation}
  \mathbf{f}^{drag} = \frac{\rho_p}{ \rho_g t_s }\mathbf{(V - U)} 
\end{equation}
where $\mathbf{V}$ is the mean weighted particle velocity at a grid point,
$t_s$ is the particle gas drag stopping time, and $\rho_p$ is the particle mass
density.  To obtain $\mathbf{V}$, a weighted sum is carried out over particles in
the eight cell volumes adjacent to each fluid grid point. Weighting functions
are used which vary inversely with the distance of the particle from the grid
point (Squires 1990, his section 5.1). Because of the periodic boundary conditions, the wake of a clump can impinge artificially on the clump from the upwind direction. To avoid this, we include the term $\mathbf{f}^{sp}$ in equation (6) as a
``sponge force'' per unit mass to restore the gas velocities downstream
of the clump at the outflow boundary plane to their initial values $\mathbf{U}_0$; this ensures constant inflow
values at the upstream boundary plane. 

The sponge force has the form
\begin{equation}
  \mathbf{f}^{sp} = \frac{1}{\tau_{sp}}(\mathbf{U} - \mathbf{U}_{0}),
\end{equation}
where $\tau_{sp}$ is some time scale. The function
$1/\tau_{sp}$ is modeled as a sigmoid in the $y$ direction very near the
outflow boundary. The parameters of the sponge were determined through test runs with the 
goals of minimizing the sponge's spatial extent and maximizing its
effectiveness without introducing numerical instabilities. Once determined, the
same values were used for all production runs. 

The (Lagrangian) equation describing the motion of particle $i$ subject to the
forces of gas drag and mutual gravity is (again neglecting the Keplerian shear term)
\begin{equation}
\frac{d\mathbf{V}_i}{dt} = 2 \mathbf {\Omega_0} \times \mathbf{V}_i + \
\mathbf{f}^{drag}_i + \mathbf{f}^{grav}_i  
\end{equation}
The term $\mathbf{f}^{drag}_i$ describing the drag force per unit mass on 
particle $i$ by the gas takes the form
\begin{equation}
\mathbf{f}^{drag}_i = \frac{1}{t_s}(\mathbf{U}(\mathbf{X}_i(t))
-\mathbf{V}_i(t)),
\end{equation}
where $\mathbf{U}(\mathbf{X}_i(t))$ is the gas velocity interpolated to the
particle's position $\mathbf{X}_i(t)$ at time $t$.

Finally, the mutual gravity force per unit mass on each particle is
\begin{equation}
\mathbf{f}_i^{grav} = G \sum_{j \neq i}^{N_p} m_j \frac{
\hat{\mathbf{X}}_{ij}(t)}{(|(\mathbf{X}_{ij}(t)| + \epsilon)^2}
\end{equation}
where $G$ is the gravitational constant, and $N_p$ is the number of particles.
$\hat{\mathbf{X}}_{ij}$  is the unit vector associated with the distance
$\mathbf{X}_{ij}(t) \equiv \mathbf{X}_i(t)-\mathbf{X}_j(t)$ between particles $j$
and $i$, and $\epsilon$ is a constant that softens the force at small
separations to prevent numerical singularities, chosen to be comparable to a
grid cell in extent.  

Eqs. (6) and (9) are solved using psuedo-spectral methods commonly used to
solve Navier-Stokes equations for a turbulent fluid (Canuto et al 1987).  
Periodic boundary conditions are assumed and the number of
nodes used in each direction is generally 128, 256, 128 with a spacing of
$2 \pi/64$. A Fast Fourier Transform (FFT) algorithm is used to evaluate the dynamical
variables $\mathbf{U}$ at the computational nodes. A second-order Runge-Kutta
scheme is used to time-advance the gas and particle velocities.    A third-order
Taylor series interpolation scheme is used to determine gas velocities at the
particle positions from values at the eight nearest neighbor nodes.    The
mutual gravity calculation is done in a brute force fashion with a $N_{p}^2$
algorithm to evaluate the force contributions from all particles.  
The code is written in Fortran 77 and is parallelized using OpenMP
directives.    The FFT calculations are done simultaneously on the planes
perpendicular to the $y$ and $z$ axes, and all loops involving particle
indices are parallelized. Typically, 3000 ``superparticles" are used, but some runs 
used more than 20000. Each ``superparticle" represents the dynamical effects, and typical response, of a large number of actual chondrules. 

Initially the particles are arranged in a uniform spherical clump.  The
initial velocities for the gas and the particles inside and outside the clump are
determined from the expressions in Nakagawa et al. (1986):
\begin{eqnarray}   
   \mathbf{U}_0 & = &  w_g \frac{\rho_p}{\rho_g + \rho_p}\left( \frac{2D\Omega}{D^2
    + \Omega^2}, \                                                              
      \frac{D^2}{D^2 + \Omega^2} - \frac{\rho_g + \rho_p}{\rho_p} , 0 \right) \\
   \mathbf{V}_0 & = & -w_g \frac{\rho_g}{\rho_g + \rho_p}\left( \frac{2D\Omega}{D^2 
    + \Omega^2}, \
      \frac{D^2}{D^2 + \Omega^2}    , 0 \right)
\end{eqnarray}
   where $D = (\rho_g + \rho_p)/\rho_g t_s$.
The initial velocity of the gas outside the clump is found by setting 
$\rho_p \rightarrow 0$  in equation (12): $\mathbf{U_0} = (0, -w_g, 0)$;
inside the clump the specific case value of $\Phi$ is used in equations (12) and 
(13) to initialize the gas and particle velocities.

Runs with single particles of different $t_s$ were made to verify that the Nakagawa et al. (1986) initial conditions are steady state solutions for the particle equations. 
The wallclock time to evolve one integration step is 0.7 sec using 256
Intel Itanium 2 processors running at 1.5 gighertz. The wallclock time to evolve 1 orbital period is 121 hours.  

\subsection{Results of the numerical model}

{\bf Figure 3} shows how, as predicted by the simple theory of section 3.3, certain combinations of clump mass loading $\Phi$
and dimension $l$ remain stable against ram pressure disruption for nebula
headwinds produced by a pressure gradient characterized by $\beta$. ${\rm We}$ is dimensionless, and the second
expression in eqn. (2) can be used to obtain the critical value of
${\rm We}_G$ which separates stable from unstable configurations of the clump, ${\rm We}_G ={\rm We}_G^*$, directly from measured values in code units (see Appendix B). We expect that ${\rm We}_G^*$ will be in the range 1-10. The combination of parameters in the stable case shown (right panels) gives ${\rm We}^*_G \sim 1$; if self gravity is turned off, the clump is disrupted in $t_{dis} \sim 1/2\pi$ orbits, whereas it will sediment into itself on a timescale $t_{sed}\sim 8$ times as long (see Appendix B). In the stable cases, a dense core is seen to continually shrink and become denser throughout the run, even as material is shed from the periphery of the clump, such that the value of $\Phi l$ for the core continues to increase (see Appendix C, {\bf figure 7}). 

\begin{figure}[t]                              
\centering                                                                   
\includegraphics[angle=0,width=3.4in,height=3.0in]{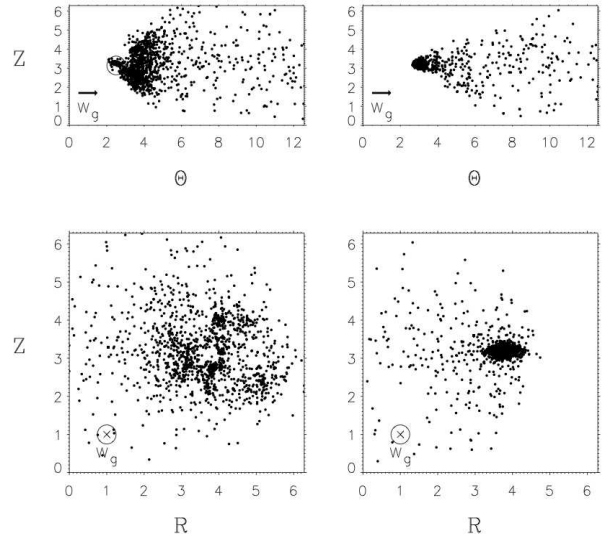}
%\includegraphics[angle=0,width=3.5in,height=3.4in]{hill_colorfigure3.ps}
%\vspace{0.1 in}
\caption{Snapshots from evolutionary models of dense clumps experiencing a
nebula headwind without (left column) and with (right column) self-gravity. Top
row: projected into a vertical plane ($z-\theta$); bottom row: projected onto
the $a-z$ plane ($a-\theta$ projections are similar), with the gas inflow into
the page. In the top left panel the open circle shows the initial size and
position of the clump. After less than an orbit, ram pressure disrupts the
gravity-free clump (left) but the gravitating clump is stable (right) and
continuing to shrink inexorably, even while being eroded. Movies showing this
evolution are available online (see Appendix C).}
\end{figure}

The numerical clump models exhibit considerable erosion over the duration of the runs from a viscously stirred surface layer, and this gradual but inexorable mass loss might lead to their disruption if arbitrarily long runs were practical at this time. However, as discussed in Appendices B and C, this large amount of erosion is a gross overestimate of what would happen in the nebula, because numerical viscosity (and consequently shear around the periphery of the clump) plays a far larger role in the current numerical models that it ever would in the actual nebula. That is, the viscously stirred and eroded surface layer in the numerical clumps contains orders of magnitude more mass than would be the case for a nebula clump of the size and density required to survive ram pressure disruption (see Appendix B). In Appendix C, we describe the movies from which {\bf figure 3} is taken and also show how the protected inner regions of the clumps are behaving exactly as predicted, both in stable and unstable regimes. For more realistic, larger clump ${\rm Re}_c$ in the nebula, a far larger fraction of the clump mass will show this behavior rather than being artificially eroded. 

The primary purpose of the numerical models is merely to provide an independent sanity check on the physics of our Weber number model and obtain some idea of ${\rm We}_G^*$, which gives us an order-of-magniture estimate for the product $\Phi l$ which nebula clumps require to survive headwinds of a particular $\beta \rho_g$. Future coding improvements are needed to allow larger (higher ${\rm Re}_c$) clumps which would more closely approach nebula conditions in terms of their balance between pressure forces and viscous forces (see Appendices B and C); these improvements will probably include an implicit time advance for the drag terms and perhaps also a tree or particle-in-mesh code for the particles.

\section{Discussion and Summary:} We have shown how self-gravity can stabilize dense clumps of mm-sized particles, which form naturally in 3D turbulence, against disruptive gas ram pressure on timescales which are sufficiently long for their constituent mineral grains to sediment towards their mutual centers and form physically cohesive ``sandpiles" of order 10-100km in size. The essence of the result is a critical ``gravitational Weber number" on the order of unity, in which self-gravity plays the role of surface tension in more familiar situations such as raindrops. Characteristic mass densities and lengthscales are determined which meet this requirement. We show numerical results which are in general agreement with the predictions of the  simple theory, for isolated, spherical initial clumps. 

The scenario we have sketched out leads from
aerodynamically size-sorted nebula particles, having the properties of
chondrules, to sizeable planetesimals formed entirely from such particles which
contain a snapshot or grab-sample of the local particle mixture. Turbulent
concentration first produces the dense zones of size-sorted  particles. The more
common, less dense of these regions may provide typical chondrule melting
environments (Cuzzi and Alexander 2006). Some of the less common, very dense zones, which we have shown elsewhere can achieve mass densities 100 times larger than that of the gas, have the potential to become planetesimals depending on their lengthscales, nebula location, and local
vorticity.  It is intriguing that our characteristic stabilized clump masses are not far from the mass inferred by Bottke et al (2005) to represent a typical primordial object in the pre-dispersal, pre-erosional asteroid belt.  

Ultimately, the sandpiles resulting from completed sedimentation will become compacted further by inevitable collisions with other sandpiles, leading to today's fairly dense
asteroids, while retaining a physical and chemical memory of their parent
particle clumps. The mechanism is easily extended to the unmelted aggregate
particles of the outer solar system, where, however, if chondrules are absent,
the size-sorting fingerprints of the process (Cuzzi et al 2001) 
might be less evident.

The conclusions of this paper differ from the suggestions in Cuzzi et al (2001), who focussed on the possible role of ultra-dense clumps with size comparable to a Kolmogorov scale (0.1 -1 km). Since that time, Hogan and Cuzzi (2007) found that particle mass loading saturates the value of $\Phi = \rho_p/\rho_g$ at a value of about 100, redirecting our attention to larger clumps in the $10^3-10^4$ km size range. Clumps and fluid structures of these large sizes are much more accessible with numerical codes, both of the standard direct simulation type and the cascade type (Appendix A). 

In this scenario, primitive bodies may not primarily represent spatial, but
perhaps temporal, samples of the particulate contents of the nebula as its
chemical, physical, and isotopic properties evolve over several Myr. The temporal, as well as the spatial, variation in the physical, chemical and isotopic properties of the concentrated particles, which are being continually altered by thermal events and mineralogical alteration in the nebula gas, can then help account for class-to-class variations between the chondrite groups. An implication of this drawn-out, inefficient process is that younger chondrite types should contain evidence of ``leftovers" or ``refugees" from earlier times, which escaped primary accretion. Several aspects of chondrite makeup are compatible with this implication. Of course, it is well known that ancient CAI minerals and less-ancient, less refractory Amoeboid Olivine Aggregates are found alongside much younger objects in the same chondrites ({\it eg.} Scott and Krot 2005). Also, the oldest chondrites (CVs) contain primarily type I chondrules (which have an age observationally indistinguishable to that of CAIs) and no (more oxidized, heavier O-isotope) type II chondrules, while the younger ordinary chondrites contain a mixture of type I and type II chondrules, some with hints of age variation even within a given chondrite (Kita et al 2005). The oldest and first-formed objects are all likely to have melted from their abundant $^{26}$Al (Kleine et al 2005, 2006; Hevey and Sanders 2006), producing differentiated achondrites and metallic objects, so it is no longer possible to dissect their primordial components. 

Of course, the real world is more complicated. Some complications are meteoritic: as only one example, CO chondrites and ordinary chondrites appear to be about the same age (younger than CV chondrites), but have very different chemical and isotopic properties (Kita et al 2005), which would, in the context of this scenario, argue for some degree of spatial (radial?) variation in the makeup of nebula particulates at that time anyway. 

On the theoretical side, of course, self-consistent numerical models which follow clump evolution in realistic turbulence with headwinds and vertical settling need to be pursued, to check these preliminary assessments which make simplistic assumptions about the local headwind and assume isolated clumps. Several recent studies have found that transient ``pressure ridges" arise in the gas in association with spiral density waves or vortical
structures (Haghighipour and Boss 2003, Rice et al 2004, Johansen et al 2007); 
in such regions, the local radial pressure gradient and headwind diminish (other regions will have unusually {\it large} headwinds). Moreover, clumps do not exist in isolation, but more realistically as a
dense core within a larger, less dense envelope which weakens the headwind felt
by the dense clump core relative to the nebula average we characterize by $\beta$. Both of
these conditions would allow some clumps to survive with smaller $\Phi l$. Other
complications include interactions between different strengthless objects,
leading to mergers or disruptions. 

Our stability criteria are most likely to be satisfied in a region fairly close to the nebula midplane ($\sim 0.01H$, see section 3.3), because at higher altitudes the vertical component of solar gravity leads to vertical settling of dense clumps at higher terminal velocities than are easily stabilized against. However, this vertical settling, even if it leads quickly to disruption of individual clumps, enhances the downward transport rate of small particles and
increases the solid/gas ratio closer to the midplane from that generally
predicted by simple 1D diffusion models ({\it eg.}, Bosse et al 2006). 

All these effects have
implications for the occurrence frequency of clumps of the requisite $\Phi l$ for
stability.  Quantitative statistical determinations of the volume
fraction of stable clumps which can become planetesimals at any given time will use, for instance, approaches such as the cascade models described in Appendix A. 
Simplified, preliminary work exploring these factors
indicates that, while the volume fraction of stable clumps is low at any given time (so the process is clearly not an efficient one), accretion rates are roughly an Earth mass per Myr in the asteroid belt region (Cuzzi et al 2007). However, considerable work remains to establish the statistical formation rate of appropriate clumps capable of following this evolutionary path and to study the evolution of clumps in realistic turbulence, with variable headwind velocity and simultaneous vertical settling. 

\acknowledgements
We thank J. Eaton, M. Gaffey, P. Goldreich, W. Hartmann, 
G. Laughlin, K. Sreenivasan, K. Squires, and A. Wray for very helpful
conversations on various aspects of this research, and A. Wray, A. 
Dobrovolskis, P. Garaud, and S. Weidenschilling for thorough reviews of the original
manuscript and a number of useful comments which have been included. N. Turner and A. Carballido also provided useful comments on an earlier version. This work was supported by a grant to JNC from NASA's Planetary
Geology and Geophysics program. Generous grants of cpu time from NASA's HEC
program were essential to the progress of this research; we thank E. Tu,
K. Schulbach, C. Niggley and R. Pesta in particular for their help along
these lines. We also thank J. Chang for coding optimization assistance.

\appendix

\section{Appendix A: Mass loading and the cascade model} 
This aspect of our work incorporates two related and important concepts: that of {\it intermittency}, and that of a statistical cascade process. For instance, it is widely known that dissipation of turbulent kinetic energy
occurs at the Kolmogorov scale; it is less widely known that the spatial
distribution of dissipation, like that of the particle concentration 
factor $C$, is highly {\it intermittent}.
Intermittent quantities are spatially and temporally unpredictable, and
fluctuate increasingly with increasing Reynolds number ${\rm Re}$. However, the 
statistical properties of intermittent quantities like $C$ (their probability distribution functions or PDFs) are
well determined on any lengthscale. In what follows, it will be necessary to
distinguish between two closely related quantities: the concentration $C$ and
the mass loading $\Phi = \rho_p/\rho_g = C \overline{\rho_p}/\rho_g$. This is
because the emergence of particle feedback on the gas due to mass loading
($\Phi$), depends both on $C$ {\it and} on the initial value of
$\overline{\rho_p}$. It is ultimately $C$ that is evolved in the cascade, but
{\it how} it evolves depends on $\Phi$. So both quantities will be alluded to
in parallel.

%\clearpage
%\begin{figure*}[t]                                 
\begin{figure}[t]                                 
\centering                                                                   
\includegraphics[angle=0,width=3.1in,height=2.5in]{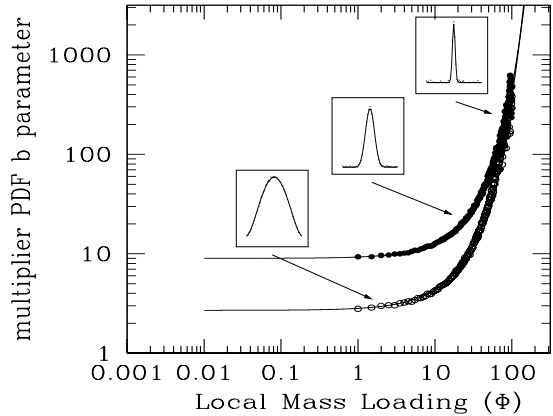}
\vspace{0.2 in}
\caption{Conditioning curves for multipliers $p(m)$; the parameter
$b$ depends on the local mass loading $\Phi$; larger $b$ indicates a
narrower PDF $p(m)$. Open symbols: for particle concentration $C$; filled
symbols: for enstrophy ($S = \omega^2$). The small insets show $p(m)$ at several values of $\Phi$. From Hogan and Cuzzi (2007).}
\end{figure}
%\end{figure*}
%\clearpage
%\begin{figure*}[t]                                 
\begin{figure}[t]                                 
\centering                                                                   
\includegraphics[angle=0,width=3.1in,height=2.4in]{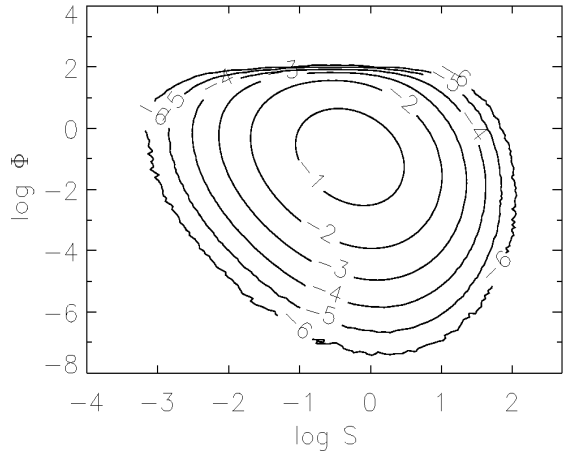}
\vspace{0.2 in}
\caption{Joint (binned) probability $\Phi \omega^2 P(\Phi,\omega^2)$ from the
 conditioned cascade model, for a 24-level case with initial 
$\overline{\rho_p}/\rho_g = 10^{-2}$. The curves of figure 1 are the $S-$averaged equivalent of this plot (and show $C$ instead of $\Phi$), but do not include the effects of mass loading. Note the flattening or
saturation at $\Phi \sim 100$, reflecting the near-vertical asymptote of the
conditioned-$b$ curve of figure 4 at $\Phi \sim 100$. A 24-level cascade is in the plausible range 
for clumps of size $\sim 10^4$ km in a nebula with $\alpha \sim 10^{-3} - 10^{-4}$.
In this figure, the enstrophy $S$ is normalized to its mean value for the 
binning lengthscale (from Hogan and Cuzzi 2007).}
\end{figure}
%\end{figure*}
%\clearpage

In its inertial range, which is extensive at high ${\rm Re}$, turbulence is a
scale-free process that is often referred to as a cascade and inspired the
approach of cascade modeling (Meneveau and Sreenivasan 1991). The
physics of transport of kinetic energy, vorticity, and dissipation from their
sources at large scales is independent of scale until viscous processes enter
at the smallest (Kolmogorov) scale. Scale-independence and ${\rm Re}-$independence
are connected because ${\rm Re}$ determines the depth of the inertial range:
${\rm Re}^{3/4} = L/\eta$ (Tennekes and Lumley 1972). Cascade models simplify this
complex, nonlinear, 3D scale-free process into a set of partition rules which
are independent of scale, or of {\it level} in the cascade. Larger ${\rm Re}$ means a
larger number of eddy spatial bifurcations (or levels) between $L$ and $\eta$,
and stronger fluctuations in intermittent properties (Meneveau and Sreenivasan
1991). Cascade models don't preserve all the information of full 3D models
(such as the tubelike spatial structures characterizing vorticity) but they do
reproduce the statistical properties (the PDFs), which are of primary
importance for our problem. The successful use of such models to describe the
intermittency of dissipation, and other quantities, in turbulence (Juneja et al
1994, Sreenivasan and Stolovitsky 1995) led us to pursue a cascade model for
turbulent concentration of particles.

A cascade model consists of a set of ``multipliers" - fractions which
define how a quantity of interest is unequally partitioned from a fluid parcel
at some level, into its equal-sized subdivisions at the next level. The PDF of
particle concentration factor $C$ is determined over all the numerous end-level
outcomes of applying a multiplier $m$ (chosen from its own PDF $p(m)$) to
each volume element as it bifurcates at each level of the cascade.  It has been 
observed that $p(m)$ is independent of level in the turbulent inertial range
(Sreenivasan and Stolovitsky 1995). Hence, multiplier PDFs determined from a
direct numerical simulation (DNS) at low ${\rm Re}$ (with a relatively small number
of levels), merely applied repeatedly over additional levels, can {\it predict}
the properties at higher ${\rm Re}$ (if nothing changes in the physics). Our cascade
model (Hogan and Cuzzi 2007) uses a common form for $p(m)$: $p(m)
\propto m^{b-1}(1-m)^{b-1}$, where the parameter $b$ determines the width of
$p(m)$ (Sreenivasan and Stolovitsky 1995).  Both the concentration $C$ and the
vorticity $\omega$ are  described by their own $b$-distributions. Multiplier PDFs with small $b (\leq 1)$ are broad, and in them multipliers much different from
$m=0.5$ occur with higher probability, producing a grainier, more intermittent
spatial distribution. PDFs with large $b (\gg 1)$ are narrow and centered on $m=0.5$; because each ``eddy bifurcation" then involves a nearly 50-50 partitioning, the resulting spatial distribution is nearly uniform and subsequent growth of $\Phi$ becomes more difficult.

The cascade model can then be used to address  the issue of how particle mass
loading can affect the cascade - that is, change the physics as the cascade
progresses. This dependence is known as {\it conditioning} of a cascade. We
found that the process can be represented quite well as two separate {\it
one-phase} conditioned cascades for $C$ and $\omega$ in which multiplier
distributions $p(m)$ for both $C$ and $\omega$ depend only on the local particle mass loading $\Phi$. We used our full 3D numerical simulations to establish how $b$ for both these properties depends on local particle or fluid properties.    

{\bf Figure 4}  shows conditioning curves for both $C$ and $\omega$ (the latter expressed as enstrophy $S = \omega^2$), as extracted from multiple 3D DNS simulations at various ${\rm Re}$ as large as 2000 (Hogan and Cuzzi 2007). Note how the multiplier PDFs extracted from the DNS results (shown using insets) get narrower (have larger $b$) as mass loading increases, choking off intermittency. In these runs, $\overline{\rho_p}/\rho_g=1$, so $C=\Phi$. These results establish the upper limit $\Phi \sim 100$ which can be obtained by turbulent concentration, regardless of the initial value of  $\overline{\rho_p}/\rho_g$. In the nebula, with a smaller initial 
$\overline{\rho_p}/\rho_g$, higher ${\rm Re}$ or deeper cascades 
(and larger ensuing $C$ values) are needed to reach this saturation point.

To determine the joint PDF of concentration $C$ and vorticity $\omega$, we
require not only the conditioned multipliers for $\omega^2$ and $C$ ({\bf Figure 4}), but also their spatial correlation. Hogan and Cuzzi (2007) found, on average, a 70-30 preference for anticorrelation at each partitioning, consistent with previous observations that particle concentration zones avoid zones of high fluid vorticity (Squires and Eaton 1990, 1991;  Eaton and Fessler 1994; Ahmed and Elghobashi 2000, 2001). Including this partitioning asymmetry factor as a weighted coin gives us a two-phase (particle-gas), conditioned cascade model which shows very good agreement with the joint PDF $P(\Phi,\omega^2)$ directly determined from our full 3D mass loaded simulations (see Hogan and Cuzzi 2007). However, to match the  full 3D DNS simulations at ${\rm Re}=2000$, the cascade model needs only about 15 levels, taking about 10 cpu-hours (for 1024 realizations) compared to over 90000 cpu hours to converge our full 3D simulation.

Using these conditioned multipliers and asymmetry factor, entire PDFs of
particle concentration (mass loading) and vorticity can be generated for
arbitrary numbers of levels (arbitrary ${\rm Re}$). For example, the 24 level model of {\bf Figure 5} (from Hogan and Cuzzi 2007) is of direct relevance for 
$10^3-10^4$ km scale dense zones of interest for the nebula. {\bf Figure 5} illustrates the saturation
of the particle concentration at $\Phi= 100$ and the fact that high $\Phi$ occurs preferentially at low enstrophy or vorticity.  The upper left quadrant is of most interest for determining the {\it numbers} of clumps which can survive to become planetesimals (see Cuzzi et al 2007 for more details). 

\section{Appendix B: Scaling between numerical code and nebula}

While the primary result of this paper - the existence of some stability regime determined by a critical gravitational Weber number ${\rm We}_G^*$ - can be expressed in a nondimensional way, some aspects of the results (specifically the numerical results and the inferred value of $\Phi l$ for the nebula) require us to pursue the relationship between nebula parameters and code units more deeply.  Values in code units are denoted by primes below. The fundamental quantities in the problem are
(1) the gas and particle densities $\rho_g$ and $\rho_p = \Phi \rho_g$;
(2) the clump diameter $l$;
(3) the local orbit frequency $\Omega(a)$ where $a$ is the distance
from the sun;
(4) the local radial pressure gradient $dP/da = dP/dx$, where $x$ is in the Hill frame;
(5) the particle stopping time $t_s$; 
(6) the velocity of material in the Hill frame ${\bf W}$, that is, relative to 
Keplerian, where ${\bf W}$ represents either the particle or gas velocity;
and
(7) the gravitational constant $G$.
The principal forces are the coriolis force due to the rotating frame
$-2\Omega_0 \times {\bf W}$, the pressure force $(-1/\rho_g)dP/dx$, and the gas
drag forces that couple the gas and particles (section 3.3). The code
unit of length is radians (the domain is 2$\pi$ radians on the short dimension 
($x$ or $z$) and the grid cell size is therefore $\Delta = 2 \pi/N$ where $N$ is the
number of grid points along that axis).  The code time unit (ctu) is
$1/\Omega'$ where $\Omega'=1$ is the code value of rotation frequency; thus the orbit period is $2 \pi$ in code time units or ctu. We treat as equivalent $a=R=x$. 

As noted in section 3.3, 
$$ \Omega(a) = (G M_{\odot} /a^3)^{1/2} \equiv (G \rho^*)^{1/2}, \hspace{0.1
in} {\rm or} \hspace{0.1 in} G=\Omega^2/\rho^*$$ 
where $\rho^* = M_{\odot}/a^3 \sim 3.8 \times 10^{-8}$ g
cm$^{-3}$ at 2.5 AU (note this is different from Safronov's (1991)
Roche density $\rho_R = 3 M_{\odot}/ 4 \pi a^3$). To normalize mass and mass
density (and ultimately establish the code value of the gravitational constant
$G'$) we assume a nebula gas mass density $\rho_o \sim 10^{-10}$ g cm$^{-3}$
(minimum mass nebula; Cuzzi et al 1993) at 2.5 AU; then code quantities
$\rho'_g = \rho_g/\rho_o = 1$ and $\rho'_p = \rho_p/\rho_o = \Phi \rho'_g = 
\Phi $, and $\rho'^* = \rho^*/\rho_o$. Also, in code units $\Omega' =
\Omega/\Omega_0 \equiv 1$ where $\Omega_o = \Omega(a_o)$ is the actual orbit
frequency at the region of interest. Then the gravitational constant in code
units, $G'$, can simply be written as $G'= \Omega'^2/\rho'^* = 1/\rho'^*$. Finally, the pressure force is written as 

$$ -{ 1\over \rho_g}{ dP \over dx }
\equiv 2 \beta a \Omega^2 = 2 \beta a G \rho^* = 2 w_g
\Omega, $$ where $\beta \approx 10^{-3}$ is typical (Nakagawa et al
1986; Cuzzi et al 1993), $G'= \rho_o/\rho^* = 3 \times 10^{-3}$. Then the dimensionless parameters $\Phi$ and $l'$ are set to satisfy
various constraints of the code (resolvable clump, adequate particle
statistics, reasonable timescale, etc) and $\beta'$ is allowed to vary, to
determine empirically the critical value of $ {\rm We}^*_G$ as the largest
value of ${\rm We}_G$ that remains stable. That is, we want $l'$ to be large
enough such that the clump is well resolved, but not so large that it fills the
entire cross section of the computational box. Typically $l'$ is only 10-20
grid cells so far, and as noted below, this exaggerates the viscous
stresses and surficial erosion of material. We want $\Phi$ to be large enough
that the clump is much denser than the gas, and we want a large enough number
of particles to act like a continuum. The value of $t_s$ should ensure that
the clump sedimentation time $t_{sed}$ is larger than its ram pressure disruption time $t_{dis}$ (see below), to provide a real test that self-gravity is preserving the clump rather than simple collapse.
 
\subsection{Important timescales in the numerical model}
The dynamical collapse time of the clump under
its own self-gravity, and in the absence of gas pressure, is $t_G = \pi
(G\rho_p)^{-1/2}$, or in code units $t'_G = \pi (G' \Phi)^{-1/2}$. The mass
loading which produces a clump dynamical time comparable to the orbit time is then
obtained using $\pi (G' \Phi)^{-1/2} = 2 \pi /\Omega' = 2 \pi$ or $\Phi =
1/4G' \sim 100$, again in fair agreement with prior expectations. However, gas
drag prevents clump collapse on this timescale, as described in section 3.1, and actual shrinkage takes a time $t_{sed}$ (eqn. 1). The number of orbits over which we need to follow the clump to ensure it really survives is roughly
$$ { t_{sed} \Omega \over 2 \pi} = { c \Omega \over 8 \pi
G \Phi  r \rho_s } = {H \Omega^2 \over 8 \pi G \Phi  r \rho_s } = {
\Omega^2\over G \rho_s}{ H\over r}{ 1 \over 8 \pi \Phi} = { \rho^* \over
\rho_s}{ H\over r}{ 1 \over 8 \pi \Phi},$$ where we have used $c = H \Omega$.
This expression can be assessed using real quantities, and is 30-300 orbits for
particle radius 300$\mu$ and mass loading $\Phi = 1000-100$. This behavior
($t_{sed} \gg 2 \pi/\Omega'$) can be controlled in the code, once the other
parameters are established as above, by adjusting the particle stopping time
$t_s$. Because of long code run times, at present we are limited to stipulating only that the 
sedimentation time significantly exceeds the nominal (self-gravity-free) disruption time; that is,
$t_{sed} \gg t_{dis}$, where
$$
t_{dis} \approx (l/w)\sqrt{2 \Phi/C_D}. 
$$
This expression for the disruption time $t_{dis}$ is derived from the ram pressure force (sect 2.3) by setting the
distance traveled in time $t$ by the windward half of the clump, under
acceleration by the ram pressure force, equal to the clump diameter (assuming
the leeward half of the clump doesn't get accelerated), and solving for $t$.
The value of $C_D$ for our clumps is on the order of unity, as we verify below.

\subsection{Numerical viscosity and the clump Reynolds number:} 
The code calls for some input viscosity $\nu'$; with
it, the Reynolds number of the clump having diameter $l'$ radians, in a
headwind of speed $w'_g$ (in code units of radians per ctu), is ${\rm Re}_c' = l' w'/2
\nu'$ where $\nu'=0.1$ is the defined code viscosity (radians$^2$/ctu). We
assume an inertial range expression within the wake of the clump to determine
the wake's Kolmogorov scale: $\eta'/l' = {\rm Re}_c'^{-3/4}$, and thus $\eta' = (2 \nu'
l'^{1/3}/w'_g)^{3/4}$ radians. Our highest resolution runs to date had a
gas relative velocity $w'_g=$ 38 radians/ctu and used a clump size $l'=2$ radians. The
nominal Kolmogorov scale associated with this flow is $\eta' =
(2(0.1)(2)^{1/3}/38)^{3/4} \sim $ 0.02 radians (compared to
the grid cell size of 0.05-0.1 radians); the wake turbulence is thus
under-resolved and the code's true viscosity is numerical: $\nu'_n \sim w'_g
\Delta$ (at the boundary of the clump).

While we are not concerned with the fine-scale details of the
wake, this marginal resolution introduces some caveats. 
The ratio of ram pressure force to viscous force (both per unit area) is, in 
general, $(\rho_g w_g^2/2) / (\rho_g \nu dw_g/dr) \sim w_g l / 4 \nu ={\rm Re}_c/2$ where
${\rm Re}_c$ is the Reynolds number of a clump, and we have approximated $dw_g/dr \sim
w_g/(l/2)$. Because we are dominated by numerical viscosity, $\nu'_n \sim w'_g
\Delta$, thus in code units ${\rm Re}_c' \sim w'_g l'/2 w'_g \Delta = l'/2\Delta
\approx 5$ in many cases so far. For comparison, the Reynolds number for {\it
nebula} clumps of interesting sizes is $> 10^6$. This means that viscous stresses around the periphery of the clump, and the 
fractional depth of the viscous boundary layer, are grossly exaggerated in the numerical model relative to the actual nebula regime of interest. The anomalously
large role of (numerical) viscosity leads to anomalously large ``erosion" from
the surface of our numerical clumps, and degrades our results in the sense that
potentially stable clumps {\it might} appear less stable, due  to the erosive
mass loss experienced over their sedimentation time. Going to higher resolution cases
(${\rm Re}'_c \sim 10$) significantly ameliorated this effect but the problem has certainly not been entirely removed.  Another way to characterize the degree of artificial erosion is to estimate the {\it physical}
Kolmogorov scale for the actual nebula/clump situation: for $\nu=10^6$ cm
sec$^{-2}$, $l=10^4$ km, and $w_g$=3800 cm/s, we find $\eta \sim$ 100m, or
$10^{-5}$ of the size of a $10^4$ km clump, compared to a fractional size of perhaps $10^{-1}$ or so in the code, as determined by numerical viscosity. Clearly, surface erosion is
a much smaller effect in the nebula case than in our crude models. While even at this resolution apparently stable configurations can be found, it would be desirable for future
studies to improve on this situation.   

{\it Clump drag coefficient $C_D$:} The drag coefficient of the clump is 
Reynolds-number-dependent and must also be scaled to nebula values to obtain an estimate of $\Phi l$.
Weidenschilling (1977) gave expressions for the drag force per unit area
(giving the pressure force per unit area), experienced by solid spheres of
different ${\rm Re}_c$ in the Stokes drag regime of interest here, as $C_D \rho_g
w_g^2/2$. For $1 < {\rm Re}_c < 800$, $C_D = 24/{\rm Re}_c^{0.6}$, or $C_D \sim 8$ for ${\rm Re}_c \sim 5$, as
in our code. The large drag coefficient and ram pressure is partly due to
low-pressure zones set up immediately behind particles in this range of ${\rm Re}_c$.
By comparison, for a solid sphere with ${\rm Re}_c > 800$ as under nebula conditions,
$C_D =0.44$. However, neither of these values might be entirely applicable to our
clumps, which are not rigid spheres. In practice, we have determined $C_D$
empirically from observed values of $t_{dis}$ with self gravity turned off.
Recalling that $t_{dis} = (l/w_g)(2 \Phi/C_D)^{1/2}$, we simply plot our
estimate of $t_{dis}$, the point at which sizeable vortex-driven internal voids
appear in the clump, against the combined parameter $(l/w_g)(2 \Phi)^{1/2}$,
giving $C_D^{-1/2}$ as the slope of the best fit straight line  ({\bf figure 6}). It
seems that $C_D \approx 1$ is a good assumption, but this calculation
should be redone with higher resolution (lower numerical viscosity) codes. At
nebula ${\rm Re}_c$, $C_D$ could reach its theoretical high-$Re$ value of 0.4 
and this is what we assume in deriving $\Phi l$.

Finally then, the several stable cases we have run ({\it eg.} figure 3) 
can be characterized by $w_g'$= 38 rad/ctu,
$\rho'^*$ = 380 (which corresponds to a minimum mass nebula 
local density $\rho_g=10^{-10}$ at
distance 2.5 AU), $\Phi = 10^3$, and $l' \approx 1$ radian, and while we have not as yet accurately determined the transition value of ${\rm We}_G$, a value ${\rm We}_G \sim 1$ is clearly stable. Solving equation (4) of section 3.3 for $\Phi l$ under
{\it nebula} conditions, {\it assuming} ${\rm We}^*_G =1$, 
taking $C_D = 0.44$ for a high-${\rm Re}_c$ situation and
assuming a nominal $\beta = 10^{-3}$, we find $\Phi l > 1.5-5 \times 10^6$,
independent of $\rho_g$ and neglecting a questionable coefficient of order
unity (because of the contribution of viscous stresses in our low-${\rm Re}_c$ code). Foreseeable refinements (specifically, lower numerical viscosity) may increase
${\rm We}_G^*$ ( the surface tension analogue gives ${\rm We}_G^* \sim 10$) and thus
decrease the value of $\Phi l$.

%\clearpage
%\begin{figure*}[t]                                 
\begin{figure}[t]                                 
\centering                                                                   
\includegraphics[angle=0,width=3.1in,height=2.8in]{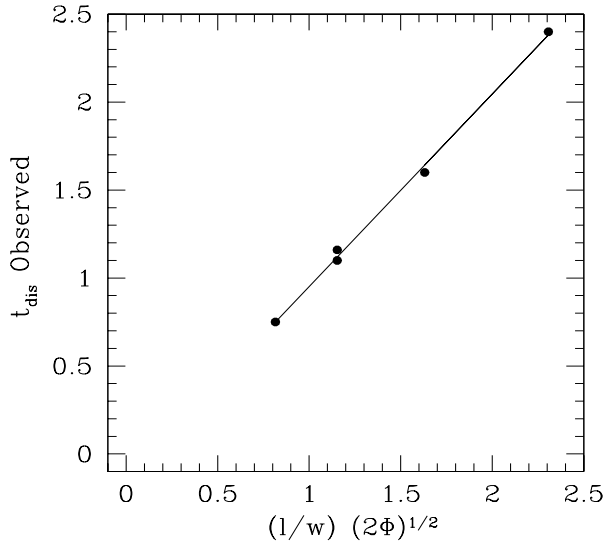}
\vspace{0.1in}
\caption{Determination of the effective drag coefficient for our clumps, from
plotting estimated disruption times $t_{dis}$ {\it vs.} the quantity $
(l/w_g)(2 \Phi)^{1/2}$; the slope is then $C_D^{-1/2}$. This determination is
slightly qualitative as to when we judge the clump to be disrupted; we decided
to use the time at which vortex-driven voids first appear in a $a-z$ planar
cross-section through the nominal center of the clump. }
\end{figure}
%\end{figure*}
%\clearpage

\section{Appendix C: Time Evolutions and Movies}
Quicktime movies showing our time evolutions are available online at 

http://spacescience.arc.nasa.gov/users/cuzzi/,

with filenames as given below. After publication, the movie files will also be 
available on the ApJ web site as mpgs. Each shows a dense clump in the Hill frame,
rotating at local Keplerian velocity, with more slowly-orbiting gas impinging
on the clump from its leading side. The two left-hand panes show the view from
along the radial axis (in the $Z-\theta$ plane; top) and along the vertical
axis (in the $a-\theta$ plane; bottom). The large right-hand pane shows the
view from the orbitally leading direction (in the $a-Z$ plane). The wake formed
by the clump is easily visible in the two left-hand panes. Cases ${\rm hill26.gif}$
and ${\rm hill29a.gif}$ are sampled at a similar time near their end-point in figure 3. File ${\rm hill26.gif}$ shows how a clump with no self-gravity is
disrupted in about 1.5 code time units for the initial conditions chosen (the
orbit period is $2 \pi$ code time units), in agreement with our simple
estimates of $t_{dis}$. Files ${\rm hill29a.gif}$ and ${\rm hill31a.gif}$ are intended to
illustrate the same (stable) Weber number case (${\rm We}_G \sim 1$) but we vary
two constituent parameters of ${\rm We}_G$ (we decrease both the gas velocity
and mass loading by a factor of three) to illustrate similarity; because of the
specific parameters chosen, ${\rm hill31a.gif}$ runs for a longer time, but is
suffering considerable erosion towards the end of the run. Nevertheless, as discussed below, it
retains a dense core that continues to contract. As discussed in Appendix B,
more realistic cases, with higher numerical resolution, would have smaller
numerical viscosity and incur less erosion. Case ${\rm hill29a.gif}$, while not run as
long as ${\rm hill31a.gif}$, appears solidly stable with a dense core that is
continuing to shrink at the end of the run. Note some interesting oscillatory
behavior, such as hinted at in our 1D compressible runs (figure 2).

\subsection{Behavior of central regions of dense clumps}

The simulations which created the movies referred to above can be used to assess the behavior of the central, dense regions of each clump in a quantitative manner. {\bf Figure 7} has three panels, showing the time variation of some properties of the particles which lie within the central regions of the clump (its densest fraction, as sampled out to some cumulative mass threshold). In the top panel, we show how much mass is being eroded from the clump overall (particles moving faster than some comoving velocity threshold are assumed to have left the clump). All three clumps are losing mass, naturally, but the non-gravitating clump 26 is losing it much more rapidly than the gravitating clumps 29a and 31a. In the central panel we show the mean mass density as measured over some fraction $f$ of the clump mass lying at the highest densities; in clumps 26 and 29a, we chose $f=0.5$, and in clump 31a, we chose $f=0.25$. In the nongravitating clump, the core density never increases and it quickly blows up, while the central regions of the gravitating clumps (removed from the artificial mass erosion at their peripheries) inexorably get denser at a steady and perhaps even slightly increasing rate (an increasing rate is predicted by the simple model). The lower panel shows the effective average radius of the region being sampled. The non-gravitating clump never shrinks at all, but both gravitating clumps do. Moreover, as the gravitating clumps shrink, their density increases faster than their linear dimension decreases, so the product $\Phi l$ increases - causing them to become {\it more} stable as time goes on. This is an argument for stability even in the presence of erosion. As expected, the density growth timescale (the sedimentation time $t_{sed}$ of equation 1) is three times smaller for the denser clump of case 29a (middle panel). Meanwhile, the apparent agreement of the variation of the fractional radius containing half of the mass seen between cases 29a and 31a in the lower panel is fortuitous, because the total masses of the clumps are changing (and eroding) at different rates and from different depths in the two cases. Nevertheless, the cores are both shrinking and becoming denser in the two stable cases. The amount of erosion that is incurred from the margins of each clump is related to (perhaps proportional to) the thickness of the viscous boundary layer relative to the radius of the clump, which can be expressed in terms of the Reynolds number. As we have argued above in this Appendix, actual nebula clumps capable of remaining stable would have Reynolds numbers five orders of magnitude larger - impossible to study with numerical models. The relevance of all viscous boundary layer effects, including the erosion that inflicts our current numerical runs, will decrease by that factor in the real situation. 

%\clearpage
%\begin{figure*}[t]                                 
\begin{figure}[t]                                 
\centering                                                                   
\includegraphics[angle=0,width=3.3in,height=3.3in]{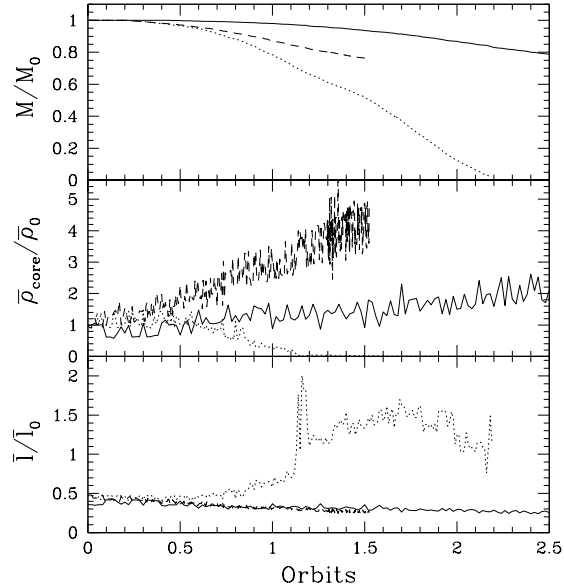}
\vspace{0.1in}
\caption{Properties of the dense central regions in three simulations, as a function of time. Cases are shown in each panel as: solid line (stable, self-gravitating case 31a); dashed line (stable, self-gravitating case 29a); and dotted line (unstable, non-gravitating case 26). Not all cases were run to the same stopping point due to resource constraints. Top: cumulative total mass retained by the clump; middle panel: average mass density of the central mass fraction $f$; bottom panel: effective radius containing mass fraction $f$. These results show that, away from the artificially perturbed perimeter regions, $We_G$-stable clumps are behaving as expected by the simple model, and $We_G$-unstable clumps do not show this behavior.}
\end{figure}
%\end{figure*}
%\clearpage

\end{document}